\documentclass[%
reprint,
superscriptaddress,
amsmath,
amssymb,
aps,
prb,
floatfix,
longbibliography
]{revtex4-1}
\usepackage{amssymb}
\usepackage{graphicx}
\usepackage{appendix}
\usepackage{color}
\usepackage{subfigure}
\usepackage[colorlinks=true,linkcolor=blue,filecolor=blue, urlcolor=blue,citecolor=blue]{hyperref}

\begin{document}
\preprint{APS/123-QED}

\graphicspath{{mainfigures/}}
\title{Tunable magnetism in bilayer transition metal dichalcogenides}
\author{Li-Ya Qiao}
\email{These authors contribute equally to the paper.}
\affiliation{Shanghai Key Laboratory of Special Artificial Microstructure Materials and Technology, School of Physics Science and engineering, Tongji University, Shanghai 200092, P.R. China}
\author{Xiu-Cai Jiang}
\email{These authors contribute equally to the paper.}
\affiliation{Shanghai Key Laboratory of Special Artificial Microstructure Materials and Technology, School of Physics Science and engineering, Tongji University, Shanghai 200092, P.R. China}
\author{Ze Ruan}
\affiliation{Shanghai Key Laboratory of Special Artificial Microstructure Materials and Technology, School of Physics Science and engineering, Tongji University, Shanghai 200092, P.R. China}
\author{Yu-Zhong Zhang}
\email[Corresponding author.]{Email: yzzhang@tongji.edu.cn}
\affiliation{Shanghai Key Laboratory of Special Artificial Microstructure Materials and Technology, School of Physics Science and engineering, Tongji University, Shanghai 200092, P.R. China}
\date{\today}

\begin{abstract}
Twist between neighboring layers and variation of interlayer distance are two extra ways to control the physical properties of stacked two-dimensional van der Waals materials without alteration of chemical compositions or application of external fields, compared to their monolayer counterparts. In this work, we explored the dependence of the magnetic states of the untwisted and twisted bilayer 1T-VX$_2$ (X = S, Se) on the interlayer distance by density functional theory calculations. We find that, while a magnetic phase transition occurs from interlayer ferromagnetism to interlayer antiferromagnetism either as a function of decreasing interlayer distance for the untwisted bilayer 1T-VX$_2$ or after twist, richer magnetic phase transitions consecutively take place for the twisted bilayer 1T-VX$_2$ as interlayer distance is gradually reduced. Besides, the critical pressures for the phase transition are greatly reduced in twisted bilayer 1T-VX$_2$ compared with the untwisted case. We derived the Heisenberg model with intralayer and interlayer exchange couplings to comprehend the emergence of various magnetic states. Our results point out an easy access towards tunable two-dimensional magnets.
\end{abstract}

\maketitle
\section{INTRODUCTION}
The breakthroughs in discovering several layered two-dimensional van der Waals magnets, for example, Cr$_2$Ge$_2$Te$_6$\cite{gong2017discovery}, CrI$_3$\cite{huang2017layer}, Fe$_3$GeTe$_2$\cite{deng2018gate}, FePS$_3$\cite{wang2016raman}, VX$_2$ (X=Se, S)\cite{bonilla2018strong,guo2017modulation}, etc., have stimulated tremendous research interest, due to their potential applications as high-performance functional nanomaterials\cite{rahman2021recent,hossain2022synthesis} and the flexible tunability via changing external controlling parameters, such as twist\cite{chen2022twist}, pressure\cite{li2019pressure},
strain\cite{ma2012evidence,son2022strain}, magnetic field\cite{wilson2021interlayer,gu2022magnetic}, electric field\cite{huang2018electrical}, doping\cite{li2017two}, stacking order\cite{zhu2021insight}, and introducing defects\cite{zhao2018surface,chua2020can}, etc.

Recently, great efforts have been devoted to tuning the magnetism of VX$_2$ using various approaches. It has been theoretically pointed out that strain can induce tunable magnetism in VX$_2$\cite{ma2012evidence,son2022strain}. Using first-principles calculations, it was shown that the magnetism will increase sharply when monolayer VS$_2$ is exfoliated from the bulk at room temperature\cite{zhang2013dimension}. Additionally, there has been a proposal for tunable magnetic phase transitions from interlayer ferromagnetic to antiferromagnetic states in bilayer VX$_2$ under pressure\cite{wang2020bethe}. Moreover, stacking order has been recognized as an effective way to tune interlayer magnetic orders in bilayer VSe$_2$\cite{li2020coupling}. Finally, it was discovered that ferroelectricity and antiferromagnetism can be coupled through a ferrovalley in bilayer VS$_2$, enabling electrically controlled magnetism\cite{liu2020magnetoelectric}. Apart from these theoretical calculations, experiments have demonstrated that Se vacancies\cite{yu2019chemically,chua2020can} can enhance the magnetism, and the deposition of Co\cite{zhang2019magnetic} or Fe\cite{vinai2020proximity} can induce the magnetism in VSe$_2$.

Despite extensive investigations, the magnetism of twisted bilayer VX$_2$ under pressure remains unexplored. It is known that twist and pressure both can tailor the properties of layered materials due to the corresponding variation of interlayer couplings\cite{wu2015interface,liu2014evolution}. As a result, many fascinating phenomena arise, such as pressure-dependent flat bands\cite{carr2018pressure,yndurain2019pressure} and band gap\cite{jiang2022tunable,gao2020band}, angle-dependent renormalized Fermi velocity\cite{bistritzer2011moire,trambly2010localization}, twist-induced site-selective phases\cite{jiang2024site}, etc. In addition to these phenomena, the magnetism of materials is greatly influenced by interlayer couplings as well. For instance, interlayer ferromagnetic and antiferromagnetic states coexist in twisted bilayer CrI$_3$, which results from a competition between the interlayer coupling and the energy cost for forming ferromagnetic-antiferromagnetic domain walls\cite{xu2022coexisting}. Therefore, it is foreseeable that magnetic phase transitions may also emerge in bilayer VX$_2$ when taking both twist and pressure into consideration.

In this paper, we investigate the tunability of magnetism with decreasing interlayer distances, which can be effectively viewed as increasing pressure,
in both untwisted and twisted bilayer 1T-VX$_2$, whose monolayer counterpart has been experimentally synthesized\cite{bonilla2018strong,sahoo2022progress}.
We find that a phase transition from ferromagnetic state (FM) to layered antiferromagnetic state (LAFM) occurs in untwisted bilayer 1T-VX$_2$
as interlayer distance decreases, where intralayer magnetic order remains ferromagnetic while interlayer becomes antiferromagnetic.
The estimated critical pressures for the phase transition of bilayer 1T-VSe$_2$ and bilayer 1T-VS$_2$ are 36.21 GPa and 42.65 GPa, respectively.
Using a simple Heisenberg model, we ascribe this phase transition to the competition between the nearest- and next-nearest-neighbor interlayer couplings,
where the former, supporting FM, plays a leading role at large interlayer distances while the latter, favoring LAFM, plays a
dominant role at small interlayer distances.
To investigate the tunability of magnetism in bilayer  1T-VX$_2$ under both twist and pressure,
we then take the case with a twisted angle of $\theta$ = $21.79^{\circ}$ as an example since it corresponds to the smallest commensurate supercell.
Although we have only investigated the smallest commensurate supercell here, it is worthwhile to mention that there exist many other twisted angles
with commensurate supercells that can be accessed with DFT as shown in the TABLE 1 of the Supplemental Material\cite{re}.
It is found that a phase transition from FM to LAFM occurs in bilayer 1T-VX$_2$ under twist solely. Remarkably, richer magnetic phase transitions are
found under pressure and the critical pressures for the phase transition are greatly reduced in twisted bilayer 1T-VX$_2$ compared with the untwisted case, where twisted bilayer 1T-VSe$_2$ undergoes phase transitions of LAFM-AFM1-AFM2 while twisted bilayer 1T-VS$_2$ experiences phase transitions of
LAFM-FM-AFM2 as interlayer distance is reduced. The critical pressures for the first and second phase transitions in twisted bilayer 1T-VSe$_2$
are 7.055 GPa and 20.73 GPa, respectively, while those in twisted bilayer 1T-VS$_2$ are 13.91 GPa and 32.21 GPa, respectively. Here, AFM1 and AFM2 are
two novel magnetic states characterized by both intralayer antiferromagnetic and interlayer ferromagnetic orders. For AFM1, the spins of overlapping
V atoms (from the top view) are antiparallel to the rests of the same layer, while for AFM2, spins of some non-overlapping V atoms (from the top view)
are parallel to that of overlapping V atoms as shown in Fig.~\ref{1D-transition}(b). By treating the non-overlapping V atoms with $C_3$ rotational symmetry
in the supercell at each layer as a pseudo-atom with a large effective magnetic moment, an effective Heisenberg model is derived to understand
these magnetic phase transitions. Please note, our aim is to extract a minimal model to comprehend tunable magnetism we discovered by DFT calculations in the twisted bilayer 1T-VX$_2$ of 21.79$^\circ$, rather than to create a unified model for all twisted angles which is out of the scope of our study. Using this model, we find that the competition and cooperation between intralayer and interlayer couplings are responsible
for the rich magnetic phase transitions in twisted bilayer 1T-VX$_2$.

The structure of our paper is organized as follows. Section~\ref{Computational-method} describes the details of the structure and the method we used.
Section~\ref{RESULTS} presents our main results, including the magnetic phase transitions and the superexchange couplings as functions of interlayer distances for both untwisted and twisted bilayer 1T-VX$_2$, the schematic diagram of magnetic phase transitions in the twisted bilayer 1T-VX$_2$ as interlayer distance decreases, the structure of the twisted bilayer 1T-VX$_2$ and its modelling. Section~\ref{DISCUSSION} presents a detailed discussion about our results, and Section~\ref{CONCLUSION} concludes with a summary.

\section{Computational method}\label{Computational-method}
Our density functional theory calculations are based on the projector-augmented-wave method\cite{blochl1994projector},
as implemented in the Vienna $ab$ $initio$ simulation package\cite{kresse1994ab,kresse1996efficiency}. We choose
the generalized gradient approximation of Perdew-Burke-Ernzerhof\cite{perdew1996generalized} as the exchange-correlation functional with the van der Waals correction of DFT-D2 developed by Grimme\cite{grimme2006semiempirical}. A plane-wave energy cutoff of 450 eV is employed. A sufficiently large vacuum distance of $20$ {\AA} is used to eliminate the interactions between periodic images of the layers in the direction perpendicular to the plane. The Brillouin-zone integration is carried out with $\Gamma$-centered K-point grids of 9$\times$9$\times1$ and 25$\times$25$\times1$ for twisted bilayer 1T-VX$_2$ and untwisted bilayer 1T-VX$_2$, respectively. The convergence criteria for the energy and atomic forces are 1$\times$$10^{-6}$eV and 0.01 eV/{\AA}, respectively. We adopt the GGA+$U$ approach of Dudarev\cite{dudarev1998electron} to include the effective on-site Coulomb interactions $U_{eff}$ between electrons of d orbitals in V atoms, where $U_{eff}=1$ eV is adopted since the fully optimized in-plane lattice constants are comparable with the experiment\cite{feng2018electronic,ji2017metallic} with $a=b=3.317${\AA} ($a=b=3.189${\AA}) for 1T-VSe$_2$ (1T-VS$_2$).
\section{RESULTS}\label{RESULTS}
\begin{figure}[htbp]
    \includegraphics[width=0.48\textwidth,height=0.45\textwidth]{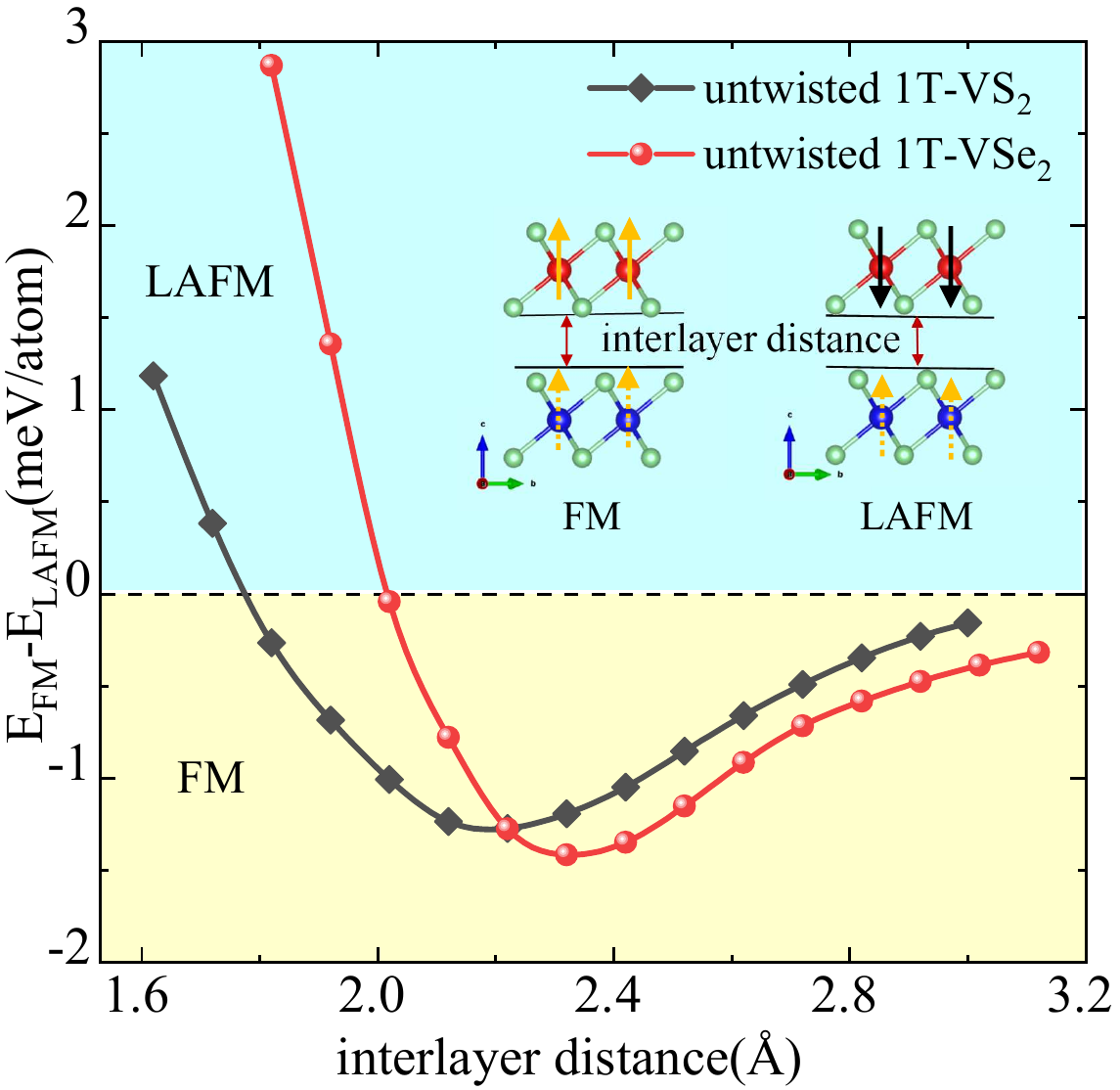}
    \caption{The energy difference between FM and LAFM $E_{FM}$-$E_{LAFM}$ as a function of the interlayer distance for untwisted 1T-VX$_2$. The inset demonstrates the specific configurations of FM and LAFM, where orange dot arrow and orange (black) line arrow denote the spin up  in the bottom layer and the spin up (down) in the top layer, respectively. }
    \label{pri-phase-transition}
\end{figure}
As pressure can tune the magnetism, we will now explore the tunability of magnetism in untwisted bilayer 1T-VX$_2$ with decreasing interlayer distances  corresponding to increasing pressure. In Fig.\ref{pri-phase-transition}, we present the evolution of energy difference between FM and LAFM as a function of the interlayer distance for untwisted 1T-VX$_2$. As can be seen, a magnetic phase transition from interlayer FM to LAFM occurs in both bilayer 1T-VSe$_2$ and bilayer 1T-VS$_2$ with the decrease of interlayer distance, which is consistent with previous study\cite{wang2020bethe}. Part of our results, such as the stability of the magnetic ground states, is double-checked by DS-PAW which is also based on the projector-augmented-wave method\cite{blochl1994projector}.

\begin{figure}[htbp]
    \includegraphics[width=0.48\textwidth,height=0.45\textwidth]{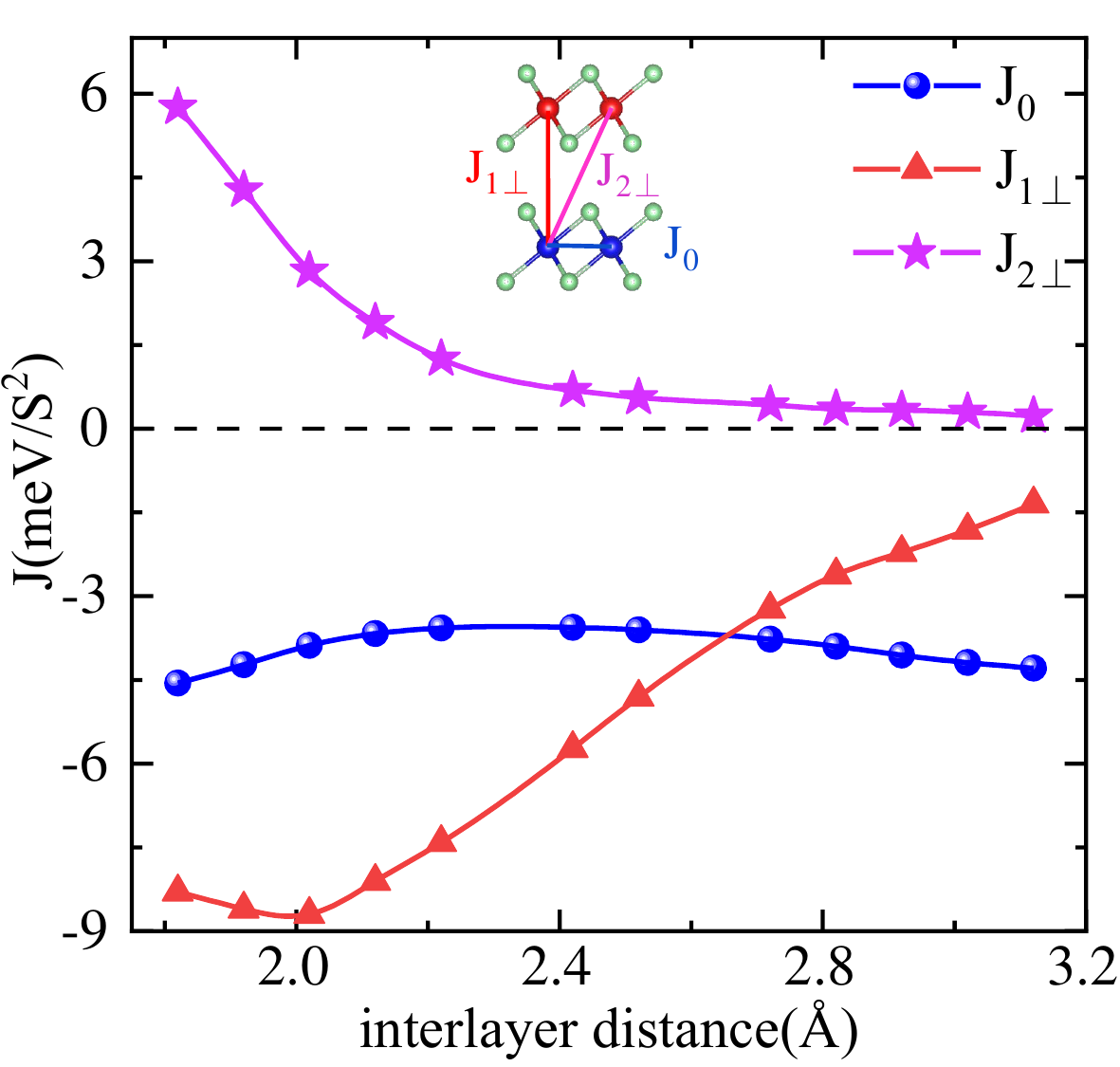}
    \caption{The superexchange couplings in untwisted 1T-VSe$_2$ as functions of the interlayer distance. The nearest-neighbor intralayer coupling $J_0$, nearest-neighbor interlayer coupling $J_{1\perp}$, and next-nearest-neighbor interlayer coupling $J_{2\perp}$ are illustrated in the inset.}
    \label{primitive-cell-Heisenberg}
\end{figure}
To gain a deeper insight into the underlying physics of this magnetic phase transition, a simple Heisenberg model $H=E_0+\sum_{ij}J_{ij}\boldsymbol{S}_i\cdot \boldsymbol{S}_j$ is employed to describe the magnetic properties of bilayer 1T-VSe$_2$, where $E_0$ is the ground state energy independent of the spin configuration and $J_{ij}$ is the superexchange coupling between local V moments of $\boldsymbol{S}_i$ and $\boldsymbol{S}_j$. Based on the energy differences among four magnetic states provided in the Supplemental Material (see Fig.S3)\cite{re}, the nearest-neighbor intralayer coupling $J_0$, nearest-neighbor interlayer coupling $J_{1\perp}$, and next-nearest-neighbor interlayer coupling $J_{2\perp}$ are derived and are summarized in Fig.\ref{primitive-cell-Heisenberg}. Obviously, $J_0$ remains ferromagnetic, leading to the intralayer ferromagnetism at all interlayer distances. In contrast, the interlayer magnetic order changes from ferromagnetic to antiferromagnetic with the decrease of interlayer distance due to the competition between $J_{1\perp}$ and $J_{2\perp}$. At small interlayer distances where $J_{1\perp}>-3J_{2\perp}$, LAFM appears. Otherwise, FM remains. Thus, compressing interlayer distances, viewed as applying pressure, can tune interlayer couplings which may induce a magnetic phase transition in untwisted bilayer 1T-VX$_2$.

\begin{figure*}[htbp]
\centering
    \includegraphics[width=0.85\textwidth,height=0.61\textwidth]{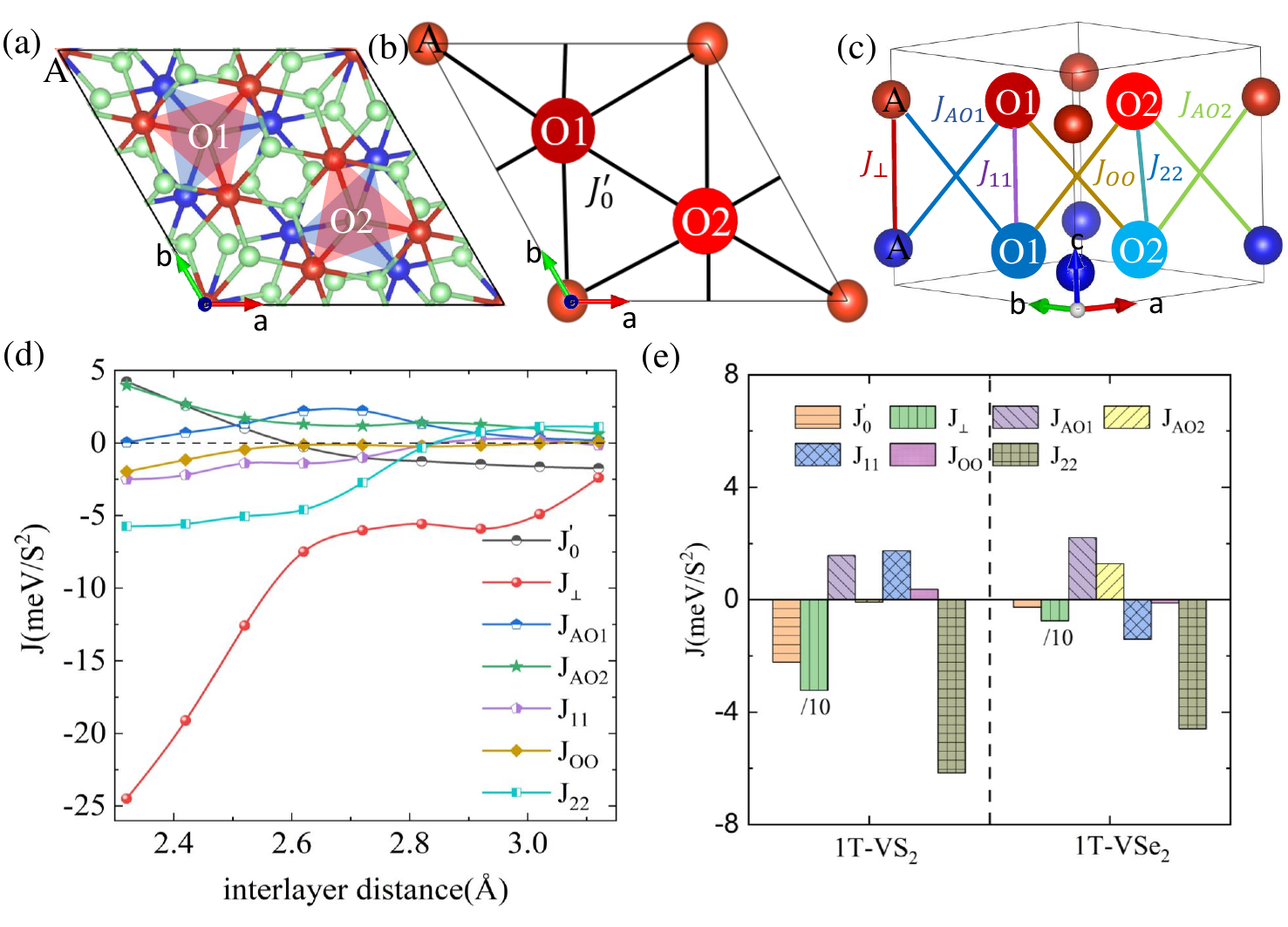}
    \caption{(a) The top view of twisted bilayer 1T-VX$_2$ with twisted angle of $\theta$ = $21.79^{\circ}$, where the red, blue, and green balls denote V atoms in the top layer, V atoms in the bottom layer, and X atoms, respectively. By treating the total magnetic moment of three non-overlap V atoms with $C_3$ rotational symmetry [locate at the vertexes of each triangle in (a)] in the supercell at each layer as an effective magnetic moment, twisted bilayer 1T-VX$_2$ can be mapped onto an effective AA stacking bilayer triangular lattice. The top view (b) and side view (c) of the effective AA stacking bilayer triangular lattice, where the intralayer nearest-neighbor coupling $J_0^{\prime}$ is shown in (b) and interlayer couplings $J_\perp$, $J_{AO1}$, $J_{AO2}$,  $J_{11}$, $J_{OO}$, and $J_{22}$ are presented in (c). (d) The superexchange couplings $J_0^{\prime}$, $J_\perp$, $J_{AO1}$, $J_{AO2}$,  $J_{11}$, $J_{OO}$, and  $J_{22}$ as functions of interlayer distance. (e) The superexchange couplings of twisted bilayer 1T-VS$_2$ and  1T-VSe$_2$ at interlayer distance of 2.32{\AA} and 2.62{\AA}, respectively}
    \label{equivalent}
\end{figure*}

Since twist can tune the interlayer couplings as well,
the twisted bilayer 1T-VX$_2$ may exhibit richer magnetic phase transitions than the untwisted one.
We then explore the tunability of magnetism in twisted bilayer 1T-VX$_2$ (X= S, Se) with decreasing interlayer distances.
The twisted angle is set to be $\theta$ = $21.79^{\circ}$ and the twisted center locates at two overlapping V atoms [see the structure in Fig.\ref{equivalent}(a)].
As the system exhibits $C_3$ rotational symmetry based on the following points: $(0,0)$, $(\boldsymbol{a}/3,2\boldsymbol{b}/3)$, and $(2\boldsymbol{a}/3,\boldsymbol{b}/3)$, we investigated 20 possible magnetic configurations
consistent with the rotational symmetry of the system as shown in Fig.S4 of the Supplemental Material\cite{re}.
Figure\ref{1D-transition}(a) presents the schematic diagram of magnetic phase transitions in
twisted bilayer 1T-VX$_2$ as a function of interlayer distance. First, it can be seen that a phase transition from FM to LAFM occurs in bilayer 1T-VX$_2$ under
twist solely (the interlayer distances of bilayer 1T-VSe$_2$ and bilayer 1T-VS$_2$ without compression are of 3.12 {\AA} and 3.0 {\AA}, respectively).
Remarkably, richer magnetic phase transitions appear in twisted bilayer 1T-VX$_2$ compared with the untwisted case as interlayer distances decrease,
where twisted bilayer 1T-VSe$_2$ undergoes phase transitions of LAFM-AFM1-AFM2 while twisted bilayer 1T-VS$_2$ experiences phase transitions of LAFM-FM-AFM2.
Here, AFM1 and AFM2 are characterized by both intralayer antiferromagnetic and interlayer ferromagnetic orders as shown in Fig.\ref{1D-transition}(b).
While, for AFM1, the spins of overlapping V atoms (from the top view) are antiparallel to the rests in the same layer, for AFM2, spins of some non-overlapping
V atoms (from the top view) are parallel to that of overlapping V atoms.

\begin{figure}[htbp]
    \includegraphics[width=0.49\textwidth,height=0.411\textwidth]{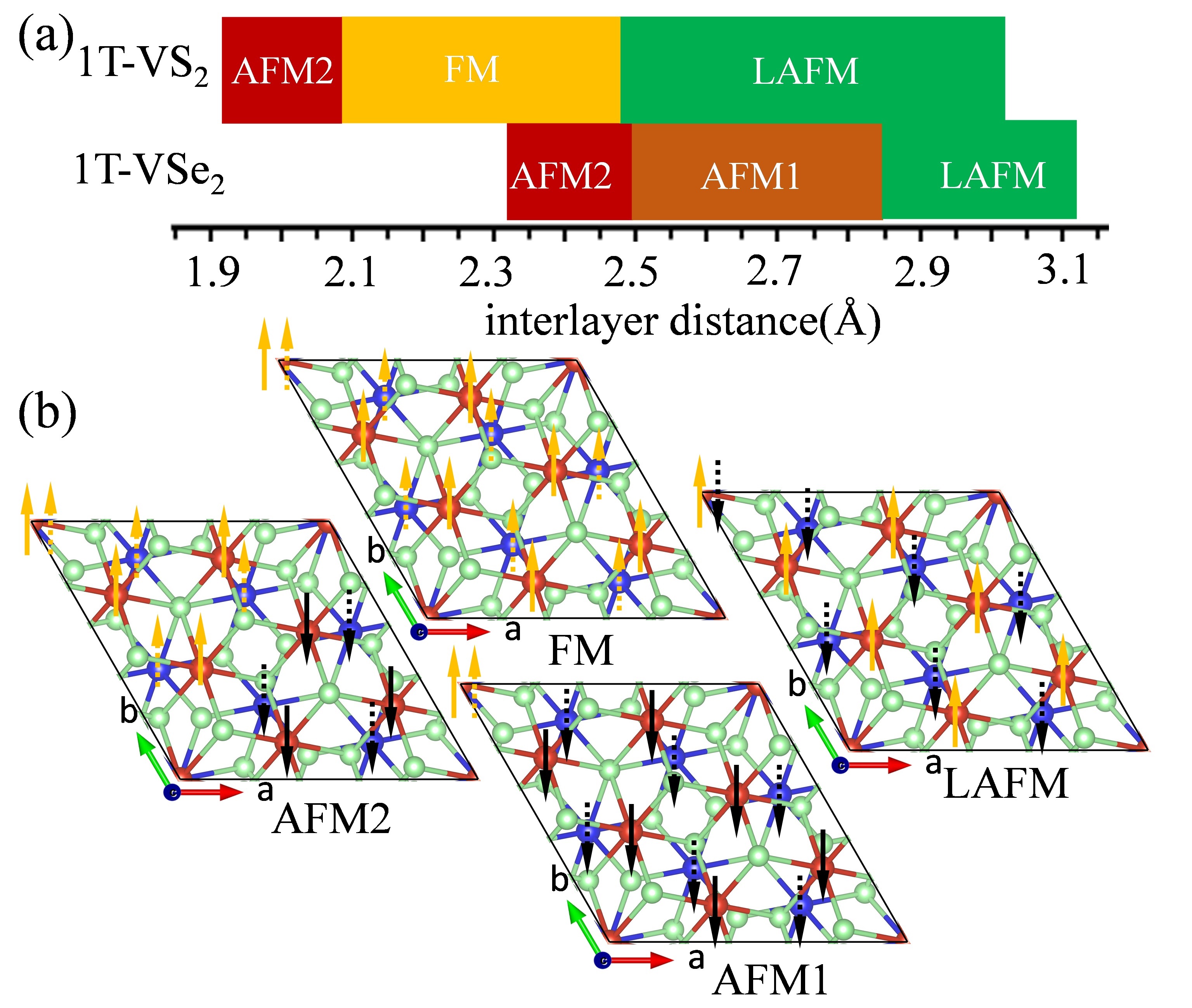}
    \caption{(a) The schematic diagram of magnetic phase transitions in the twisted bilayer 1T-VS$_2$ and 1T-VSe$_2$ as functions of interlayer distance. (b) The specific configurations of LAFM, FM, AFM1, AFM2, where orange (black) dot arrow and orange (black) line arrow denote the spin up (down) in the bottom layer and the spin up (down) in the top layer, respectively.}
    \label{1D-transition}
\end{figure}
To investigate the underlying physics of the magnetic phase transitions in twisted bilayer 1T-VX$_2$,
we will now map twisted bilayer 1T-VX$_2$ onto an effective lattice and establish an effective Heisenberg model based on this lattice.
According to the $C_3$ rotational symmetry of the system, we identify three distinct regions of V atoms within the supercell at each layer,
namely regions A, O1, and O2, where both O1 and O2 include three equivalent V atoms located at the vertexes of each triangle as shown in Fig.\ref{equivalent}(a).
Besides, the magnetic moment on each V atom remains approximately 1$\mu B$ irrespective of magnetic orders.
Thus, we can treat the total magnetic moment of V atoms in region O1 (O2) within the supercell at each layer as an effective magnetic moment that is three times
larger than the original moment of each V atom. Ultimately, the twisted bilayer 1T-VX$_2$ is mapped onto an effective AA stacking bilayer triangular lattice which
includes three inequivalent sublattices at each layer, namely A, O1, and O2, as presented in Fig.\ref{equivalent}(b) and \ref{equivalent}(c).
By taking into account the nearest-neighbor intralayer coupling, nearest-neighbor interlayer couplings, and next-nearest-neighbor interlayer couplings,
we obtain an effective Heisenberg model for this effective AA stacking bilayer triangular lattice that
consists of $J_0^{\prime}$, $J_\perp$, $J_{AO1}$, $J_{AO2}$,  $J_{11}$, $J_{OO}$, and $J_{22}$ as shown in Fig.\ref{equivalent}(b) and \ref{equivalent}(c).
Please note that, by carefully examining the relationship between the Heisenberg model of the original lattice and the Heisenberg model of the effective bilayer triangular lattice, it can be found that, although the A, O1, and O2 sites are inequivalent, the nearest-neighbor intralayer couplings between any two of them are equal. In addition, based on the fact that both the magnetic state of monolayer VX$_2$ and the magnetic phase transition of untwisted bilayer 1T-VX$_2$ can be well understood by merely involving the nearest-neighbor intralayer exchange coupling $J_0$ and twist between layers will not change the paths for spin exchanges between two intralayer vanadium atoms, it is reasonable to infer that inclusion of intralayer exchange coupling up to nearest neighbor is sufficient for modelling the twisted bilayer 1T-VX$_2$.
Thus, we use only one intralayer parameter $J_0^{\prime}$ in our effective Heisenberg model.

We proceed to employ this minimal effective Heisenberg model to understand the underlying physics of the phase transitions in twisted bilayer 1T-VX$_2$.
Since we study 20 magnetic configurations whose energy differences are provided in the Supplemental Material (see Fig.S5)\cite{re},
but the effective Heisenberg model has only 7 coupling parameters, we utilized a weighted least squares method, which has also been
used to determine the Heisenberg model in iron-based superconductors\cite{glasbrenner2015effect}, to determine these parameters,
where the rules for the weights are as follows: (1) Since the magnetic ground states exhibit interlayer symmetry, the weight of the magnetic configurations with interlayer symmetry should be greater than
that of the magnetic configurations without interlayer symmetry. (2) For magnetic configurations with interlayer symmetry,
the weight of a configuration with lower energy should be greater than that of a configuration with higher energy, and similarly for those without interlayer symmetry. The specific formulas satisfying above rules for the weights are given in the Supplemental Material [Eq.(4) and Eq.(5)]\cite{re}. These weights ensure the error between the lowest total energies obtained from DFT calculations and predicted from above Heisenberg model less than 0.1 meV/atom. Then, the 7 couplings can be readily derived.

Figure \ref{equivalent}(d) presents the 7 couplings of the aforementioned minimal effective Heisenberg model for twisted bilayer 1T-VSe$_2$
as functions of interlayer distance. As can be seen, at large interlayer distances, intralayer coupling $J_0^{\prime}$ maintains ferromagnetic,
meanwhile, all of the interlayer couplings except for $J_\perp$ exhibit antiferromagnetic behavior ($J_{OO}$ is negligibly small).
Consequently, the system favors an intralayer ferromagnetic and interlayer antiferromagnetic ground state, namely, LAFM. In contrast,
at intermediate interlayer distances, the vertically interlayer couplings, including $J_\perp$, $J_{11}$, and $J_{22}$, exhibit ferromagnetic characteristics,
which will favor an interlayer ferromagnetic order. Meanwhile, the diagonal interlayer couplings, including $J_{AO1}$ and $J_{AO2}$,
maintain antiferromagnetic properties ($J_{OO}$ is negligibly small), which suggest an interlayer antiferromagnetic order between A and other sublattices,
corresponding to AFM1. At small interlayer distances, due to $J_{11}$ and $J_{22}$ maintaining interlayer ferromagnetism, and $J_0^{\prime}$
being intralayer antiferromagnetic, the magnetic orders on the O1 and O2 sublattices are interlayer ferromagnetic and intralayer antiferromagnetic.
Simultaneously, the strong ferromagnetic $J_\perp$ and strong antiferromagnetic $J_{AO2}$ result in an interlayer ferromagnetic order between overlapping
V atoms (A sublattices) with an intralayer antiferromagnetic order between A and O2, namely, AFM2.

Although, twisted bilayer 1T-VS$_2$ has the same magnetic ground states as twisted bilayer 1T-VSe$_2$ at large and small interlayer distances,
the ground state at intermediate distances is different. To understand this discrepancy,
the couplings of the twisted bilayer 1T-VS$_2$ and  1T-VSe$_2$ at interlayer distance of 2.32{\AA} and 2.62{\AA}, respectively, is presented in Fig.~\ref{equivalent}(e).
Obviously, the ferromagnetic intralayer coupling of $J_0^{\prime}$ still plays a dominant role, which is different from the case of twisted bilayer 1T-VSe$_2$,
resulting in intralayer ferromagnetism. Meanwhile, both $J_\perp$ and $J_{22}$ exhibit strong ferromagnetic couplings, leading to interlayer ferromagnetism.
Therefore, FM is the ground state at intermediate distances for the twisted bilayer 1T-VS$_2$.

\section{DISCUSSION}\label{DISCUSSION}
Here, by systematically investigating the tunability of the magnetic properties in untwisted and twisted bilayer 1T-VX$_2$ (X = S, Se),
we find that, the twisted system is easier to access richer phase transitions than untwisted one,
indicating that the combination of twist and pressure provides an efficient way to tune the magnetism of layered materials.
Although some studies have reported the presence of charge density wave in
layered 1T-VSe$_2$\cite{feng2018electronic,fumega2019absence,coelho2019charge} and 1T-VS$_2$\cite{kim2021electronic,van2021full},
it does not undermine the reliability of the magnetic phase transitions in twisted bilayer 1T-VX$_2$(X = S, Se) demonstrated in this paper
because twist can narrow the bandwidth of twisted layered materials, thereby facilitating the emergence of magnetism.
For example, while bilayer graphene itself lacks of magnetism, ferromagnetism emerges in the twisted bilayer graphene\cite{sharpe2019emergent,chen2020tunable}.

We noticed that, while the detected magnetic phase transition in untwisted bilayer 1T-VX$_2$ was attributed to the competition between interlayer Pauli and Coulomb repulsions, which favor interlayer antiferromagnetism with short interlayer distance, and the kinetic energy, which supports interlayer ferromagnetism with large interlayer distance\cite{wang2020bethe}, the magnetic phase transitions in twisted bilayer 1T-VX$_2$ seem to challenge this theoretical framework. We have shown that an interlayer ferromagnetic state occurs at small interlayer distances which is corresponding to strong interlayer Pauli repulsion, while an interlayer antiferromagnetic state emerges at large interlayer distances which refers to weak interlayer Pauli repulsion.

In the untwisted bilayer 1T-VX$_2$, the intralayer nearest-neighbor coupling $J_0$ remains nearly constant as the interlayer distance decreases,
and the magnetic phase transition of the system arises from the competition between interlayer nearest-neighbor and next-nearest-neighbor couplings. However, in the case of twisted bilayer 1T-VX$_2$, the intralayer nearest-neighbor coupling $J_0^{\prime}$ gradually
changes from ferromagnetic to antiferromagnetic, playing a significant role in the magnetic phase transitions. The different behavior in intralayer nearest-neighbor hopping between untwisted and twisted cases is due to the presence of directly overlapping X atoms (from the top view) between the two X sublayers that are located within the two V sublayers after twist, resulting in the enhancement of the interactions between these overlapping X atoms,
which can ultimately affects the superexchange process between intralayer V atoms. Noticeably, there are infinite numbers of twisted angles that can result in commensurate structures with presences of directly overlapping X atoms between the two X sublayers that are located within the two V sublayers after twist. The twisted angle and corresponding number of atoms in the commensurate supercell are shown in Fig.S 6 of the Supplemental Material\cite{re}. We argue that the twisted bilayer 1T-VX$_2$ which exhibits fruitful magnetic phases may be common in nature.

The twist angle plays a crucial role in tuning the properties of materials.
For instance, while untwisted bilayer CrI$_3$ exhibits an interlayer antiferromagnetic ground state, interlayer ferromagnetic and antiferromagnetic states
coexist in twisted bilayer CrI$_3$ with twisted angle of $\theta\leq3^\circ$
\cite{xu2022coexisting}.
At last, left-twisted ($\theta=32.2^{\circ}$) and right-twisted ($\theta=27.8^{\circ}$) bilayer 2H-VSe$_2$ show
distinct responses to an electric field\cite{shen2021exotic}.

Since the structure of twisted bilayer 1T-VX$_2$ holds $C_3$ symmetry, our focus here is primarily on magnetic
ordered configurations consistent with $C_3$ symmetry.
Please note, we can not rule out possibility of existence of other magnetic ground states without restriction of symmetry,
since one can not exhaust all antiferromagnetic configurations in the thermodynamic limit. However,
even if it might happen, our conclusion, i.e. twist and pressure able to induce rich magnetic phase transitions, remains unchanged.

\section{CONCLUSION}\label{CONCLUSION}
In conclusion, we systematically investigate the magnetic phase transitions of untwisted and twisted bilayer 1T-VX$_2$ and reveal
the underlying physics based on corresponding Heisenberg model. For untwisted bilayer 1T-VX$_2$, a phase transition from FM to LAFM is found as interlayer distance decreases, which is attributed to competition between the nearest- and next-nearest-neighbor interlayer couplings. In contrast, richer magnetic phase transitions are found under pressure and the critical pressures for the phase transition are greatly reduced in twisted bilayer 1T-VX$_2$ compared with the untwisted case, where twisted bilayer 1T-VSe$_2$ undergoes phase transitions of LAFM-AFM1-AFM2 while twisted bilayer 1T-VS$_2$ experiences phase transitions of LAFM-FM-AFM2. It is necessary to mention that a phase transition from FM to LAFM occurs in bilayer 1T-VX$_2$ under twist solely. By employing an effective Heisenberg model, we find that the competition and cooperation between intralayer and interlayer couplings are responsible for the rich magnetic phase transitions in twisted bilayer 1T-VX$_2$. Our results provide guidances in exploring magnetic properties of twisted layered materials.
\section{ACKNOWLEDGEMENT}
This work is supported by National Natural Science Foundation of China (NOs.12004283, 12274324) and Shanghai Science and technology program(No.21JC405700).
We gratefully acknowledge HZWTECH for providing computation facilities.
\bibliography{VSe2_reference}

\end{document}